\documentclass[twocolumn,prb,aps,floatfix,showpacs,superscriptaddress]{revtex4}
\usepackage{amsmath,amsfonts,epsf,natbib,color,subfloat}
\usepackage{graphicx}% Include figure files
\usepackage{graphics}
\usepackage{subfigure}
\usepackage{epstopdf}
\usepackage{dcolumn}% Align table columns on decimal point
\usepackage{bm}% bold math
\usepackage[mathlines]{lineno}% Enable numbering of text and display math
\usepackage[colorlinks,breaklinks,linkcolor=red,citecolor=blue,urlcolor=blue]{hyperref}

\begin{document}
%\bibliographystyle{plainnat}
%\draft
%\topmargin 0.1cm

\title{Constructing optimal local pseudopotentials from first principles}

\author{Wenhui Mi}
%email{mwh@calypso.cn}
\author{Shoutao Zhang}
\author{Yanming Ma} 
\email{mym@calypso.cn}
\affiliation{State Key Laboratory of Superhard Materials, Jilin University, Changchun 130012, China}
\affiliation{Beijing Computational Science Research Center, Beijing 100086, P. R. China}
\author{Maosheng Miao}
\email{miaoms@gmail.com}
\affiliation{Materials Research Lab. University of California Santa Barbara, CA 93110}
\affiliation{Beijing Computational Science Research Center, Beijing 100086, P. R. China}
\date{\today}

\begin{abstract}
Local pseudopotential (LPP) is an important component of the orbital free density functional theory (OF-DFT), which is a promising large scale simulation method that can still maintain information of electron state in materials. Up to date,  LPP is usually extracted from the solid state DFT calculations. It is unclear how to assess its transferability while applying to a much different chemical environment. Here we reveal a fundamental relation between the first principles norm-conserving PP (NCPP) and the LPP. Using the optimized effective potential method developed for exchange functional, we demonstrate that the LPP can be constructed optimally from the NCPP for a large number of elements. Our theory also reveals that the existence of an LPP is intrinsic to the elements, irrespective to the parameters used for the construction. Our method provides a unified method in constructing and assessing LPP in the framework of first principles pseudopotentials.  
\end{abstract}


\maketitle

\section{introduction}
In recent years, orbital-free density functional theory (OF-DFT)\cite{WangCarter2000,wesolowski2013} has attracted increasing interests due to its capability of simulating thousands to millions of atoms.\cite{Shin2012,Vora2010,Ligneres2007,Zhouj2005,Bhatt2005,Jiang2004,Gonzalez2001,Pearson1993}  Comparing with other large scale simulation methods, OF-DFT maintains the electronic structure information and is potentially applicable for systems undergo large chemical changes. 
However, OF-DFT has not become a mainstream method for large-scale simulation, due to the lack of both an accurate kinetic energy density functionals (KEDFs) and local pseudopotentials (LPPs) that are highly transferable in different chemical environments. The recently developed nonlocal kinetic energy functional showed promising results for both metals and semiconductors.\cite{Huang2010,xia2012} On the other hand, it is still unclear how to proceed in systematically constructing transferable LPPs. Although several families of empirical LPPs are already available,\cite{Goodwin1990,Fiolhais1995,Fiolhais1996,Watson1998,Jesson2000,Wang2003,Zhou2004,Zhou2005,Huang2008} most of them are constructed to fit the solid state DFT results and only work for a small variation of chemical environment. 

One important concept in developing first principles pseudopotentials is the norm-conserving (NC) condition,\cite{Hamann1979, Troullier1991} 
\begin{equation}
\int_{\Omega }n_{i}^{\text{NCPP}}(r) d^{3}r
=\int_{\Omega }n_{i}^{\text{AE}}(r)d^{3}r,
\end{equation}
in which $n_{i}^{\text{NCPP}}(r)$ and $n_{i}^{\text{AE}}(r)$ represent the orbital densities of NC pseudopotential and all-electron solutions, and $i$ is the angular momentum number. This condition requires the pseudo-charge enclosed within the core radius for each orbital must be identical to that of all-electron results and therefore ensures the transferability since it conserve the change of wavefunctions versus energy:\cite{Hamann1979, Troullier1991}
 \begin{align}
-\frac{1}{2}\frac{\partial}{\partial \varepsilon}\frac{\partial}{\partial r}\ln R&(r,\varepsilon)|_{\varepsilon=\varepsilon_{i},r=r_{c}}\nonumber \\
&=\frac{1}{r_{c}^{2}R^{2}(r_{c},\varepsilon_{i})}\int_{0}^{r_{c}}R^{2}(r,\varepsilon_{i})r^{2}dr,
\end{align}
in which $R(r,\varepsilon)$ is the radial wavefunction of angular momentum $l$. The thus constructed NCPPs are different for different angular momentum channels (orbitals) and therefore are non-local. Apparently, the NCPPs can not be used in OF-DFT calculations since the electrons in an orbital free scheme should feel the same potential. It is a long time question whether LPPs can be constructed while still satisfying the NC condition. Here, we try to tackle this problem by altering it into the following question: how can we optimally construct LPPs that are orbital independent from the NCPPs that are orbital dependent.                                                                                                         

We would like to point out the similarity between constructing LPPs and obtaining the local exchange functional from its explicit Hartree-Fock form. 
As defined in Hartree-Forck method, the exact exchange energy is the functional of single particle wave functions, {\it i.e.} 
\begin{equation}
E_{x}^{\textrm{exact}}=-\frac{1}{2}\sum\limits_{i,j}\int d^{3}r_{1} d^{3}r_{2} \frac{\phi^{\ast}_{i}(r_{1})\phi^{\ast}_{j}(r_{2})\phi_{i}(r_{2})\phi_{j}(r_{1})}{|r_{1}-r_{2}|}. 
\end{equation}
For systems without spin-orbit interaction, spin freedom is trivial to add and therefore is omitted throughout the paper. 
The corresponding exchange potential is non-local and orbital dependent. 
\begin{equation}
v_{x}^{i}\phi_{i}(r_{1})= -\sum\limits_{j\neq i}\int d^{3}r_{2} \frac {\phi^{\ast}_{j}(r_{2})\phi_{j}(r_{1})}{|r_{1}-r_{2}|} \phi_{i}(r_{2})
\end{equation}
Slater proposed a density average of Hartree-Fock nonlocal orbital dependent potentials which is known as Slater potential.\cite{Slater1951}
\begin{equation}
v^{\textrm{Sl}}(r_{1}) = \frac{\sum\limits_{i}\phi^{\ast}_{i}(r_{1})v_{x}^{i}\phi_{i}(r_{1})}{\sum\limits_{i}\phi^{\ast}_{i}(r_{1})\phi_{i}(r_{1})}
\end{equation}
Afterwards, a method that construct the optimal local exchange potential variationally from the explicit orbital expression of the exchange energy was established and applied for many cases.\cite{KLI1995,Grabo1998,Bylander1995}
\begin{align}
V_{x}^{\textrm{OEP}}(\bm{r})&=\frac{\delta E_{x}[\{\phi_{i}\}]}{\delta\rho(\bm{r})}\nonumber\\
&=\sum_{i=1}^{N}\int d^{3}r'\frac{\delta E_{x}[\{\phi_{i,\tau}\}]}{\delta\phi_{i}(\bm{r'})}\frac{\delta \phi_{i}(\bm{r}')}{\delta \rho(\bm{r})}+c.c.
\end{align}
 This optimized effective potential (OEP) method can be transformed into an Algebraic equation, 
\begin{align}
V_{x}^{\textrm{KLI}}(\bm{r})&=\frac{1}{2\rho(\bm{r})}\sum_{i=1}^{N}|\phi_{i}(\bm{r})|^{2}\{u_{xi}(\bm{r}) \nonumber\\
&-\frac{1}{|\phi_{i}(\bm{r})|^{2}}\nabla\cdot(\tilde{\psi}^{\ast}_{i}(r)\nabla\phi_{i}(r))   \nonumber\\
&+(\overline{V}^{\textrm{KLI}}_{xi}-\overline{u}_{xi})\}+c.c. \label{KLI}
\end{align} 
It can be simplified to an algebraic Krieger-Li-Iafrate construction(KLI)\cite{KLI1995} by neglecting the second term $-\frac{1}{|\phi_{i}(\bm{r})|^{2}}\nabla\cdot(\tilde{\psi}^{\ast}_{i}(r)\nabla\phi_{i}(r))$. The Slater potential is the first and the major term of it. It is worth to notice that OEP can be derived directly from the fact that the OEP orbitals are the first order shifts of the HF orbitals while conserving the charge density.\cite{Miao2000,Kummel2003}
 
We will show in this paper that the same idea of optimally constructing local and orbital-independent exchange potentials from the exact orbital dependent exchange potential can be used to construct local pseudopotentials. The essential issue is the conservation of the NC condition, which as shown by our work, can be satisfied for large number of elements in the periodic table. We will also show that while NC condition is preserved, the Slater potential will be identical to KLI potential, which is an excellent approximation of the exact OEP of the NC pseudopotentials. On the other hand, our work also reveals that the transferable LPPs can not be constructed for many elements, despite whichever scheme is used. Rather, as we demonstrated, the existence of a highly transferable LPP is an intrinsic property of an element.

\section{optimized effective pseudopotential}
We assume first principles NC pseudopotentials $v_{i}(r)$ for an atom, which are different to each angular momentum channel. Correspondingly, the total ion-electron interaction energy can be written as 
\begin{equation}
E_{i-e}=\sum_{i}\int n_{i}(r) v_{i}(r)d^{3}r. 
\end{equation}
The expected local pseudopotential should reproduce 
the above total energy for different electron configurations of an atom except a constant shift. 
Such local potential can be calculated by taking a density derivative of 
$E_{i-e}$, {\it i.e.} 
\begin{equation}
v(r)=\delta E_{i-e}/\delta n(r). 
\end{equation}
Applying the OEP scheme and the KLI 
approximation originally developed for HF exchange potentials, 
one can obtain the corresponding KLI equation\cite{KLI1995,Grabo1998} as:
\begin{equation}
v(r) =\sum_{i}\frac{n_{i}(r)}{n(r)}%
v_{i}(r) +\sum_{i}\frac{n_{i}(r) }{n(r)}%
(\bar{v}^{i}-\bar{v}_{i}^{i}) ,  \label{KLI}
\end{equation}
where the quantities with a bar above them are the orbital averages of the
potentials, $v(r) $ and $v_{i}(r)$, {\it i.e.} 
$\bar{v}^{i} = \int{n_{i}(r)v(r)} d^{3}r$, 
$\bar{v}_{i}^{i} = \int{n_{i}(r)v_{i}(r)} d^{3}r$, 
 and the summation of $i$ goes through all the angular 
momentum channels. Again, the spin freedom is not considered and the spin 
index is omitted here. 
The first term is the long range Slater averaged 
potential, $v_{\text{Slater}}$ and the second term is a short range correction. 
The derivation is similar to that of KLI equation 
for local exchange potential.\cite{KLI1995,Grabo1998} It relies on the fact 
that the wavefunctions of local NC pseudopotential are the 
first order perturbation of the wavefunctions of semi-local NC pseudopotential.

The above KLI construction will change the potential 
outside the core because of the non-Slater term. 
This is undesirable because the pseudopotentials outside the core should be identical to the 
true ionic potential. One possible solution is to construct KLI potential 
only inside the core and keep the original ionic potential outside the 
core. However, the resulted potential will be discontinuous at the core 
boundary. On the other hand, it is not hard to notice that the Slater 
potential that is the first and the major term of KLI potential does not 
change the potential outside the core region. This is because the first principles NC pseudopotentials outside the core are
identical for different channels, and the Slater average is a density average 
of all angular momentum channel.

\begin {table*}
\begin{ruledtabular}
\caption{Constructing and testing OEPP of Ga with various atomic configurations, eigenvalues of  pseudo-atom and core radius, and with and without nonlinear core correction. The defininition of $\rho^{c}_{s}$, $\rho^{c}_{p}$ and $\delta_{\rho}$ can be found in Appendix~\ref{appB} \label{improve_Ga} }.  
\begin{tabular}{ccccrrrrrr}
 Methods  & Configuration & $R_{nlcc}(a.u)$ &$R_{cut}\left(s\ /p\right)(a.u)$&$\epsilon_{s}(eV)$  &$\epsilon_{p}(eV)$ &$E_{tot}^{ps}(eV)$&$\rho^{c}_{s}$&$\rho^{c}_{p}$&$\delta_{\rho}$ \\
\hline
TM-NCPP	  &$4s^{2}4p^{1}$				&--	&2.75\ /2.75	&-9.1750	&-2.7384	&-58.0460	&0.7200 	&0.3924 	&0.0000\\	
OEPP	                                &$4s^{2}4p^{1}$		&-- 	&2.75\ /2.75	&-9.1966 	&-2.7500 	&-58.1033 	&0.7250 	&0.3807 	&0.0182\\
OEPP$(\delta=0.0018 a.u)$	&$4s^{2}4p^{1}$		&-- 	&2.75\ /2.75	&-9.1791 	&-2.7414 	&-58.0189 	&0.7244 	&0.3789 	&0.0244\\
\hline	
TM-NCPP	                        &$4s^{2}4p^{1}$		&1.75 	&2.75\ /2.75	&-9.1750 	&-2.7384 	&-104.8345 	&0.7200 	&0.3924 	&0.0000\\	
OEPP	                                &$4s^{2}4p^{1} $		&1.75 	&2.75\ /2.75	&-9.1944 	&-2.7512 	&-104.8916 	&0.7248 	&0.3808 	&0.0172\\
\hline			
TM-NCPP      				&$4s^{0.5}4p^{2.5}$	&-- 	&2.2\ /2.2	&-10.7706 	&-3.8364 	&-47.9618 	&0.5085 	&0.2480 	&0.0000 \\	
OEPP					&$4s^{0.5}4p^{2.5}$	&-- 	&2.2\ /2.2	&-10.8364 	&-3.8418 	&-48.0362 	&0.5290 	&0.2451 	&0.0100\\
\hline
TM-NCPP				&$4s^{0.5}4p^{2.5}$	&1.75 	&2.2\ /2.2	&-10.7706 	&-3.8364 	&-93.3344 	&0.5085 	&0.2479 	&0.0000\\
OEPP					&$4s^{0.5}4p^{2.5}$	& 1.75 	&2.2\ /2.2  	&-10.8292  	& -3.8421 	&-93.4084 	&0.5285 	&0.2451 	&0.0094 \\		
\end{tabular}
\end{ruledtabular}
\end{table*}

\begin{table*}
\begin{ruledtabular}
\caption{$s$ and $p$ energy levels and the excitation energies for several
atomic configurations of Mg, Ga and Sb, calculated by all electron potentials, Troullier-Martins pseudopotentials, OEPPS as well as the BLPS pseudopotentials. The two Troller-Martins pseudopotentials are constructed at the two configurations for OEPP and BLPS, respectively. 
Units are in eV. \label{trasferability}}
\begin{tabular}{cccrrrrr}
&  &  & AE & TM$_{\textrm{OEPP}} $&TM$_{\textrm{BLPS}}$&OEPP& BLPS \\ \hline
Mg & &  &$3s^{2}3p^{0}$ &$3s^{1}3p^{1}$&$3s^{2}3p^{0}$&$3s^{1}3p^{1}$&$3s^{2}3p^{0}$\\ 
\\
& $s^{1}p^{1}$ & $s$	 & -5.7701  & -5.7702 	&-5.7391	&-5.9200&-5.8323		\\
&  & $p$ 	    			&-2.1295    & -2.1295  	&-2.1268	&-2.0873&-2.1230		\\
&  & $\Delta E$			& 3.5233    & 3.5367 	& 3.5105	&3.6926&3.5876		\\
& $s^{2}p^{0}$ & $s$	 	& -4.7878  & -4.8104	&-4.7878	&-4.8216&-4.8058 	\\
&  & $p$ 				& -1.3773  & -1.3760	&-1.3773	& -1.2668&-1.3382		\\
&  & $\Delta E$ 			& 0.0000   & 0.0000	& 0.0000	&0.0000		&0.0000\\
& $s^{1}p^{0}$ & $s$ 		&-11.5278 & -11.4939	&-11.4416&-11.5977&-11.4980	\\
& & $p$		              		& -7.1642  &-7.1456	&-7.1272&-6.9529&-7.0234		\\
& & $\Delta E$ 	     		&8.0742	  &8.0824	&8.0459&8.1341&8.0802		\\
\hline
Ga &&  &$4s^{2}4p^{1}$ &$4s^{2}4p^{1}$&$4s^{2}4p^{1}$&$4s^{2}4p^{1}$&$4s^{2}4p^{1}$\\
& &&$-$&$r_{c}=2.75$&$-$&$\delta=0.0018$&$-$\\
\\
&$s^{1}p^{2}$	&$s$	&-10.2808  &-10.2781  &-10.2511  &-10.3780&-10.0103     \\
& 			       &$p$	&-3.5000	   &-3.5027    &-3.5015     &-3.5368&-3.4351    	 \\
&	            &$\Delta E$	&6.6124	   &6.6118     &6.5993     &6.6461&6.4143       	\\
&$s^{2}p^{1}$     &$s$		&-9.1750	   &-9.1750   &-9.1750	&-9.1791 &-8.9113     	\\
&			       &$p$	&-2.7384	   &-2.7384   &-2.7384    &-2.7414&-2.6713     	\\
&		    &$\Delta E$	&0.0000	   &0.0000    &0.0000	 &0.0000	     &0.0000	\\
&$s^{1}p^{1}$     &$s$		&-17.7538 &-17.7185 &-17.6771  &-17.8063&-17.4160     \\
&		   	      &$p$		&-10.2007 &-10.1656 &-10.1639  &-10.1323&-10.0634      \\
&    		   &$\Delta E$		&13.3385  &13.3279  &13.1192    &13.3720&13.0427        \\	
\hline
Sb&&  &$5s^{2}5p^{3}$ &$5s^{1}5p^{4}$&$5s^{2}5p^{3}$&$5s^{1}5p^{4}$&$5s^{2}5p^{3}$\\
\\
& $s^{1}p^{4}$ & $s$	 &-13.8933  & -13.8933 	&-13.8761	&-13.3325&-12.2457	\\
&  & $p$ 	    			&-5.5668     & -5.5668  	&-5.5657 	&-5.6422&-5.4695		\\
&  & $\Delta E$			& 8.2094     &8.2021 	& 8.2027	        &7.5444&6.6853		\\
& $s^{2}p^{3}$ & $s$	 	& -13.0893  & -13.0731	&-13.0893	&-12.4615&-11.5606 	\\
&  & $p$ 				& -4.9991    & -4.9998	&-4.9991	        & -5.0675&-4.9692		\\
&  & $\Delta E$ 			& 0.0000     & 0.0000	& 0.0000	       &0.0000		&0.0000\\
& $s^{1}p^{3}$ & $s$ 		&-21.7887  & -21.7689	&-21.7467      &-21.2557&-19.9793	\\
& & $p$		              		& -12.8751 &-12.8494	&-12.8542      &-12.9483&-12.6818	\\
& & $\Delta E$ 	     		&17.3400   &17.3237	&17.3259       &16.7555&15.6744		\\
\end{tabular}
\end{ruledtabular}
\end{table*}

The further consideration to the above problem relies on the relation between the NC condition 
and the OEP and KLI equations, which will be examined here. 
Denoting the orbital densities calculated from local pseudopotential and
first principles NC pseudopotential as 
$n_{i}^{\text{LPP}}(r)$ and $n_{i}^{\text{NCPP}}(r)$, the preservation of NC condition requires
\begin{equation}
\int_{\Omega }n_{i}^{\text{LPP}}(r) d^{3}r
=\int_{\Omega }n_{i}^{\text{NCPP}}(r)d^{3}r,
\label{nc}
\end{equation}
where $\Omega $ denotes the space inside the core radius. 
Considering that the change of the potential will make a first order 
perturbation to the wavefunctions, one can prove 
the following relation for the potentials if the NC condition Eqn.\ref{nc} 
is satisfied (see Appendix A for proof),
\begin{equation}
\int_{\Omega }n_{i}^{\text{LPP}}(r)(v(r)-v_{i}(r))d^{3}r=0.
\label{nv}
\end{equation}
Eqn.\ref{nv} can also be written as $\bar{v}^{i}-\bar{v}_{i}^{i}=0$. 
It suggests that if the NC condition is 
retained during the construction of local pseudopotential, 
the second term of the KLI 
potential will become zero, indicating that the Slater potential is 
identical to KLI potential. 

KLI potential is an approximation of exact OEP potential by neglecting a term involving 
the diversity of orbital change.\cite{KLI1995,Grabo1998} 
$\sum\limits_{i}\nabla\cdot(\tilde{\psi}^{\ast}_{i}(r)\nabla\phi_{i}(r))$, in which 
$\tilde{\psi}_{i}=\psi_{i}-\phi_{i}$ and $\psi_{i}$ and $\phi_{i}$ are the wavefunctions of the original NC pseudopotentials and 
the constructed local pseudopotentials.
This simplification can be interpreted as a mean-field approximation, since the neglected terms averaged over 
the ground state density $\rho(r)$ is zero, or equally the integration of the neglected term over space vanish, {\it i.e.}
\begin{equation}
I=\int d^{3}r \sum\limits_{i} \nabla\cdot(\tilde{\psi}^{\ast}_{i}(r)\nabla\phi_{i}¨)=0
\end{equation}
If the construction of local pseudopotential retains the NC condition, 
this integral should be 0 for each individual orbital. 
\begin{align}
I_{i}&=\int_{\Omega} d^{3}r \nabla\cdot(\tilde{\psi}^{\ast}_{i}(r)\nabla\phi_{i}(r)) \nonumber\\
&=\oint_{\Omega}dS \tilde{\psi}^{\ast}_{i}(s)\nabla\phi_{i}(s)\cdot\hat{n} = 0
\end{align}
While the NC condition Eqn.\ref{nc} is satisfied, the local pseudopotential will generate pseudo-wavefunctions that are 
identical to the wavefunctions of all electron potentials and the NC pseudopotentials outside 
the core region, which means $\tilde{\psi}_{i}$ and $\nabla \tilde{\psi}_{i}$ should be 
exactly 0 at the core boundary and in the region out of the core. Therefore, we have shown that 
an optimized local pseudopotential can be constructed by taking the Slater 
average of the NC pseudopotentials if and only if the NC condition is kept during the construction. 
If the NC condition can be retained, the local Slater averaged pseudopotential is identical to 
the KLI potential of the semi-local NC pseudopotentials, which is very close to the exact OEP.

%%%%%%%%%%%%%%%%%%%
\section{Calculation details}
%%%%%%%%%%%%%%%%%%%

The FHI98 code\cite{FHI98PP}is modified to generate and test the proposed pseudopotentials. For comparison, the TM-NCPP\cite{Troullier1991} potentials are also generated by using FHI98 code. The details of LPP construction for a set of 27 elements, including the atomic configuration, the cutoff radius etc. are listed in Table S1. In order to improve the transferability, the nonlinear core-valence exchange-correlation\cite{nlcc} is included for some of the elements. The OEPP cutoff radii are adjusted to minimize the values of  $\delta_{l}$ , $\delta_{\rho}$ and $\delta_{E_{l}}$ (the definition of these quantities is given below ). The Kleinman-Bylander\cite{KB}  form of pseudopotentials are used in solid calculations and we truncates the angular momentum at $l_{max}=2$. 

The structural relaxations and bulk property calculations were carried out using KS-DFT as implemented in the CASTEP code.  \cite{CASTEP}  Both local density approximation (LDA)\cite{ldaca} exchange-correlation functionals and generalized gradient approximation (GGA) in Perdew-Burke-Ernzerhof (PBE) \cite{pbe} form are used for exchange-correlation functionals. The appropriate energy cutoff and Monkhorst-Pack $k$ meshes were chosen to ensure that enthalpy calculations for each system were well converged to less then 0.5meV/atom. A Fermi-Dirac smearing with a width of 0.1 eV is used for all metal systems. 

OF-DFT test calculations are carried out using our recently developed ATLAS software,\cite{ATLAS} which is based on an efficient real space finite-difference method. We employ the Wang-Govind-Cater KEDF\cite{WGC99} with the parameters: $\gamma =2.7$, $\alpha=(5+\sqrt{5})/6$ and $\beta=(5-\sqrt{5})/6$. These parameters have been demonstrated to work well for Mg and Al systems.\cite{WGC99,Huang2008}  In finite-difference OF-DFT calculations, the order of finite-difference ($N$) and the real space mesh gap $h$ (determine the max plane-wave cutoff energy $E_{cut}$ in the reciprocal space $E_{cut}=\pi^{2}/2h^{2}$) should be tested for different systems. In our calculations for Mg and Al systems, both energies converge to less than 0.1meV/atom, while setting $N=4$ and $h= 0.2 \AA$. 

%%%%%%%%%%%%%%%%%%%%
\section{Test results}
%%%%%%%%%%%%%%%%%%%%

\subsection{Atom level}

In order to check whether a local NC pseudopotential can be constructed for various elements, 
we define $\delta_{l}=\int_{\Omega }n_{l}(r)
(v_{l}(r)-v_{\text{Slater}}(r))d^{3}r$ as a
measurement of the deviation from NC condition for the Slater averaged pseudopotential. 
According to Eqns. (\ref{nc}) and (\ref{nv}), if  $\delta_{l}=0$ 
then the NC condition is retained. Furthermore, if $\delta_{l}$ is 
small, the changes of the orbitals are also small.  $\delta_{l}$ 
estimates the energy changes for each orbital, {\it i.e.} 
$\varepsilon_{i}^{\text{LPP}}-\varepsilon _{i}^{\text{NCPP}}\approx\delta_{l}$, therefore they 
are good measures for the accuracy of LPPs. 
\cite{Nagy1997} 

In the current work, we are
going to focus on  main group 
elements that only contain $s$ and $p$ electrons in their valence.  
For some p-block elements such as Ga and In, the d levels are close to the valence. Although they are 
completely filled, they may have effects to chemical bonds, and are included in the valence in many first principles 
pseudopotential constructions. However, the omission of them in the valence usually will not yield 
errors that exceeds the use of OF-DFT and the local pseudopotentials. 
Therefore, we keep the $d$ electrons in the core for all $s$ and $p$ elements. 

\begin{figure}
\includegraphics[width=8.5cm]{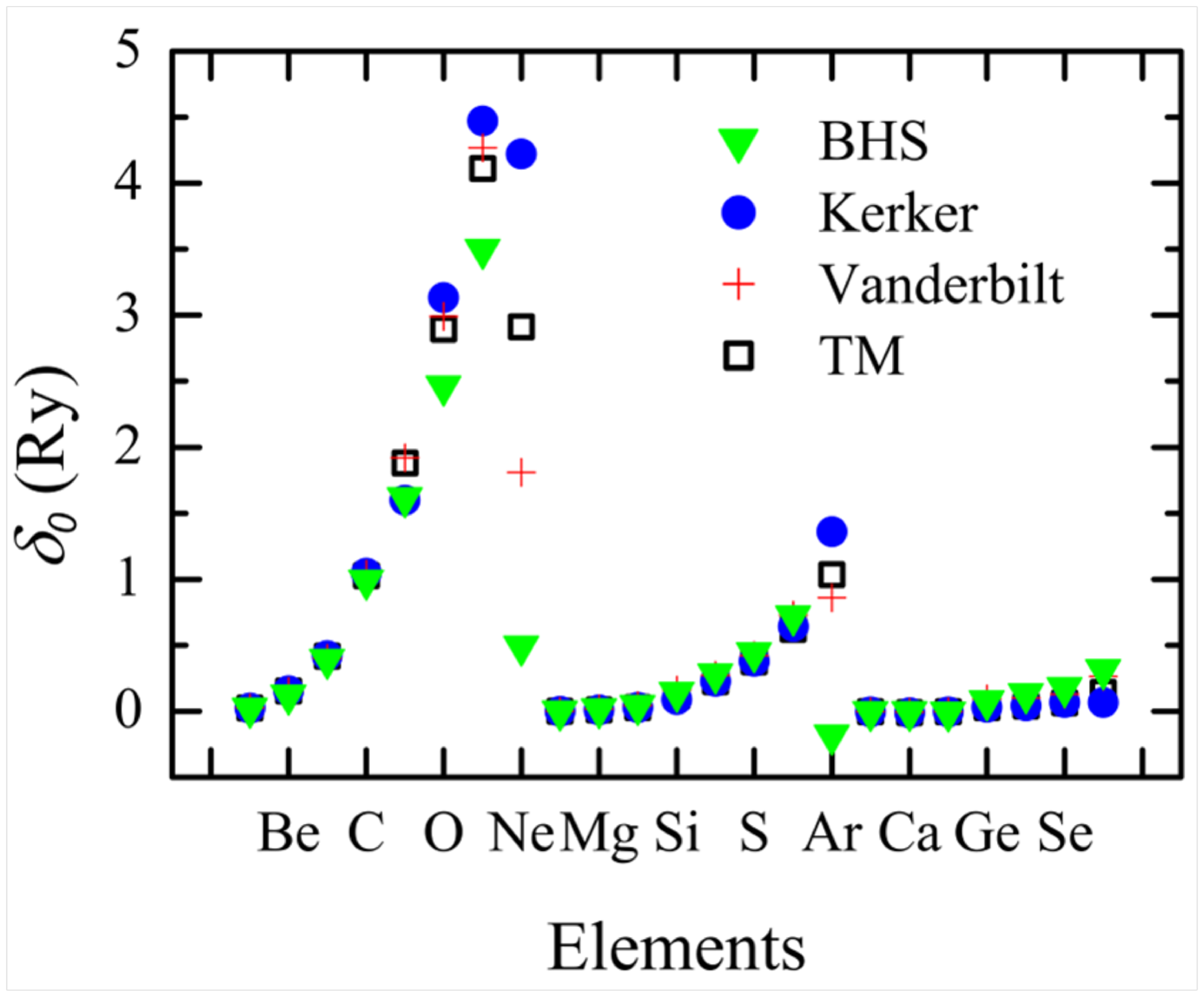}
\caption{\footnotesize (Color figures available online.) The NC deviation values $\delta_{0}$ for elements from Li to Br. The original pseudopotentials are constructed by Troullier-Martins (TM), Kerker,Vanderbilt and BHS methods, and are represented by open squares, green triangles, red crosses and blue circles, respectively. \label{delt}}
\end{figure}

In the $sp$ systems, we need to use only one 
parameter because $\delta_{0}=-\delta_{1}$ (NC deviation value). 
Figure \ref{delt} shows $\delta_{0}$ for
elements from Li to Br. The Slater potentials are constructed from four
kinds of NC pseudopotentials including 
BHS,\cite{Bachelet1982} 
Kerker, \cite{Kerker1980} Vanderbilt\cite{Vanderbilt1985} and
TM.\cite{Troullier1991} In general, 
$\delta_{0}$ becomes smaller with increasing atomic number. However, for 
the atoms in the same row in periodic table, the higher the atomic number, 
the smaller the $\delta_{0}$ value. For most of the elements, $\delta_{0}$ 
does not depend on the type of pseudopotentials except the noble gas atoms. 
This indicates the capability of constructing LPP is intrinsic to elements and 
do not depend on the method of construction.  

Although OEPP is constructed directly from NCPP through the Slater average, we still have freedoms while constructing the NCPPs, which we can utilize to optimize the performance of OEPP. These include the choice of the atomic configuration, the core radius, the use of nonlinear core correction (NLCC)\cite{nlcc} etc. Furthermore, since the Slater average might change the atomic orbital energies of the NCPP, we do not need to keep the all electron orbital energies as the NCPP energies during its construction. Thus, we are free to adjust the original atomic orbital energies in NCPP construction in order to optimize OEPP.  We choose Ga as an example to show the effect of above settings and adjustments on the improvement of OEPP accuracy. The testing results are shown in Table \ref{improve_Ga}, including the eigenvalue of each orbital $\epsilon_{s}$ and $\epsilon_{p}$, the pseudo-atom total energy $E_{tot}$, the integrated densities of $s$ and $p$ orbitals inside the core radii ($\rho^{c}_{s}$ and $\rho^{c}_{p}$), and the differences of integrated total density inside core radii for NLPP and OEPP ($\delta_{\rho}$). For these values, the smaller the differences between OEPP and the corresponding NCPP, the more accurate the OEPP is. As shown in Table \ref{improve_Ga}, we find for elements like Ga that has very small  $\delta_{0}$, OEPP works well. It reproduces the total energy and the orbital energies at the constructed atomic configurations. The variation of the constructing configuration and the other parameters such as cutoff radii does not significantly affect its accuracy. 

Furthermore, we examine the transferability of the OEPPs by calculating the orbital energies and the total energies of atomic configurations other than the one at which the NCPP and OEPP are constructed. The results of selected elements including Mg. Ga and Sb are shown in Table \ref{trasferability}. The configurations that pseudopotentials are constructed are shown in the first row for each element. The total energies are referred to the the ground-state energies obtained by the same pseudopotential. The results are also compared with the bulk-derived local pseudopotentials (BLPS).\cite{BLPS} As shown in Table \ref{trasferability}, we find that both OEPP and BLPS work fairly well for Mg and Ga. In contrast, OEPP orbital energies and total energies show large improvements than BLPS for Sb. Our test results reveal that the transferability of OEPPs are better than BLPS for the elements later then Ga. 

We extend our study to other elements in the periodic table. In general, we find OEPP works well for large number of elements including almost all s-block elements and many p-block elements. For many of them, for example Mg, the accuracy and transferability of OEPP is as good as NCPP. The important features of OEPP and its corresponding NCPP for a number of elements are listed and compared in Table S1. The OEPP features are also compared with available BLPS in Table S2.  As shown in these tables, the performance of OEPP and BLPS are in general similar, showing Slater average as a favorable method of constructing LPP directly from electronic structure of atoms. Furthermore, OEPP outperforms BLPS in many cases, especially for elements later than In.

\begin{figure}
\includegraphics[width=8.5cm]{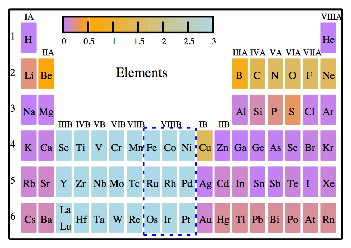}
\caption{\footnotesize (Color figures available online)
%\scalebox{2.2}{\includegraphics{a.eps}}
Color map on the feasibility of constructing highly transferable local pseudopotentials for the elements in the periodic table. The color represents the value of $\delta_{\rho}$ as shown by the color bar. The values of $\delta_{\rho}$ for transition metals and rare-earth metals are set at 3. \label{Elements}}
\end{figure} 
\begin{figure} 
%\scalebox{1.2}{\includegraphics{ppal.eps}}
\includegraphics[width=8.5cm]{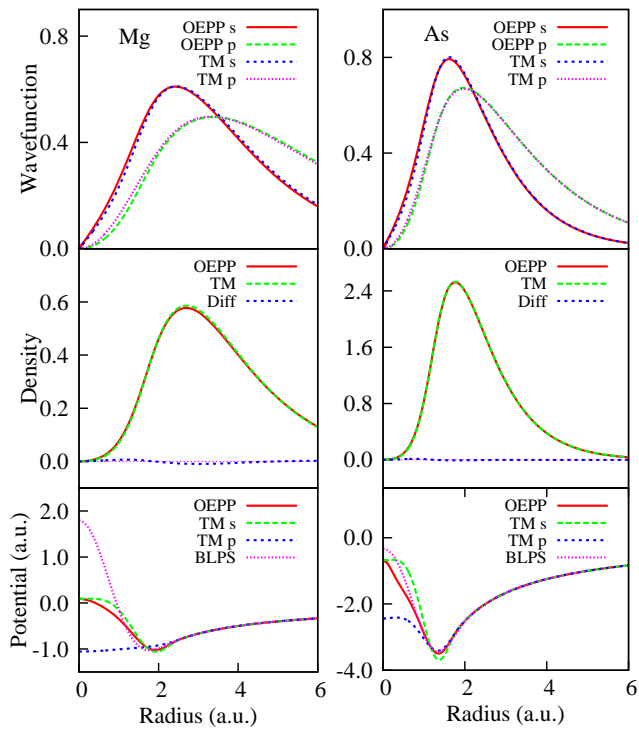}
\caption{\footnotesize (Color figures available online) The radial wavefunction of $s$ and $p$ orbitals, the total density and the potential profile for Mg (left panels) and As (right panels), obtained using the Trouliers-Martins pseudopotentials (TM), BLPS and OEPP. \label{wavefunctions}}
\end{figure}

An important point elucidated by the OEPP construction and test calculations is that the existence of LPP is intrinsic to an element. Despite the use of different parameters, atomic configurations and types of NCPPs, the resulting LPPs do not perform well for certain elements if the NC condition is not preserved. We can use the deviation from the NC condition, {\it i.e.} $\delta_{\rho}=\int_{\Omega}\ |n_{\text{OEPP}}(r)-n_{\text{NCPP}}(r)\ |d^{3}r$ of an element as a good measure for the possibility of constructing LPP with high transferability. This condition is weaker than the NC condition, which require the preservation of charge density inside the cutoff radius for each angular momentum channel. Therefore this condition is necessary but not sufficient. We also find $\delta_{\rho}$ is not very sensitive to different construction of NCPPs. We present them as a color map in the periodic table (Fig. \ref{Elements}). For those elements colored in purple, the LPPs with high transferability might be constructed. The elements for which LPPs are difficult to construct include second row elements, all transition metals, f-block metals and some late p-block elements. $\delta_{\rho}$ values for transition metals and rare earth metals actually go beyond the largest value in the range and are therefore fixed at 3. 

An immediate question is why OEPP can work for many elements, considering the large change of the potentials in each angular momentum channel after making the Slater average. We present the wavefunctions, the total densities and the potentials obtained by using NCPP and LPP in Fig. \ref{wavefunctions} for Mg and As. It shows clearly that despite the large variation of the potential, the wavefunctions of NCPP and OEPP are almost identical. This is the direct result of the fact that the pseudo-wavefunctions are constructed under many constrains, including nodeless, identical to full electron wavefunction out of core region and the norm-conserving condition. While comparing with the BLPS, we find a large difference between the two LPPs especially for the region that is close to the nucleus. 

\subsection{Bulk properties}

\begin{figure}
\includegraphics[width=8.5cm]{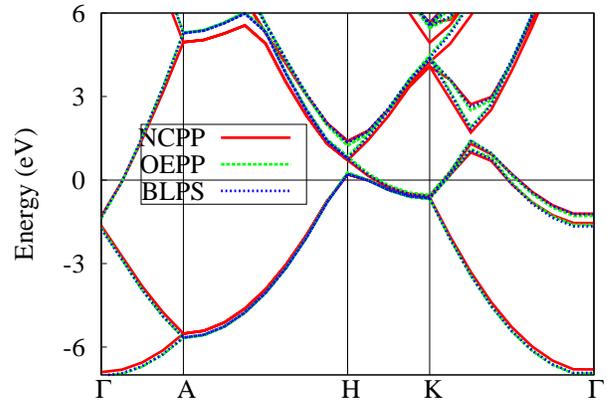}
\caption{\footnotesize (Color figures available online.) Band structure of hcp Mg obtained by use of TM-NCPP(Red), OEPP(Green) and BLPS(Blue) respectively.
\label{Mg_band}}
\end{figure}

\begin{figure}
\includegraphics[width=8.5cm]{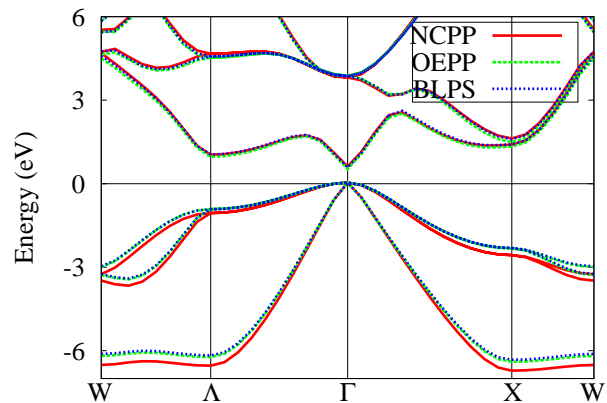}
\caption{\footnotesize (Color figures available online.) Band structure of GaAs in ZB structure obtained by use of TM-NCPP(Red), OEPP(Green) and BLPS(Blue) respectively.
\label{GaAs_band}}
\end{figure}

We now test the accuracy of OEPPs in real materials. Using Mg and GaAs as examples, we first calculate the band structures. Despite the very different way of construction, the OEPP and BLPS yield very similar band structures for both Mg (Fig.\ref{Mg_band}) and GaAs (Fig. \ref{GaAs_band}). The results also compare well with the NCPP band structures, indicating the good accuracy of both LPPs. The OEPP bands are closer to BLPS than to NCPP results, although both local pseudopotentials are constructed by totally different methods. 

\begin{figure}[!h]
\includegraphics[width=8.5cm]{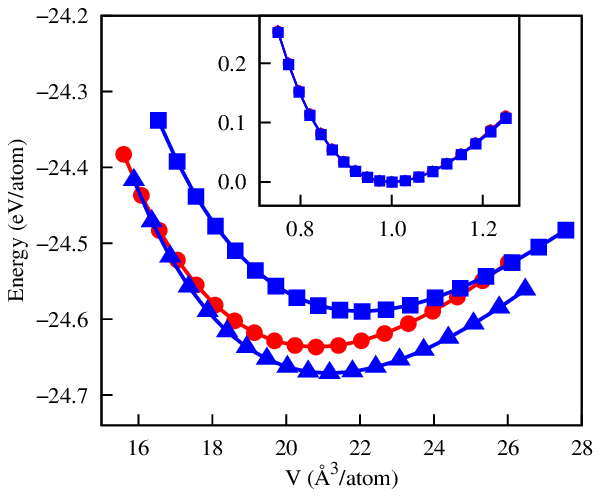}
%\scalebox{1.4}{\includegraphics{Mg_hcp_eos.eps}}
\caption{\footnotesize (Color figures available online.) The equation of states (EOS) of hcp Mg, using KS-DFT with LDA exchange functional and three different pseudopotentials, including 
Trouliers-Martins pseudopotential TM-NCPP (square), OEPP (circle) and BLPS (triangle). The inset show the EOS of total energies shifted by the equilibrium total energy as functions of the atomic volumes scaled by the equilibrium atomic volume.  
\label{Mg_eos}}
\end{figure}

\begin{figure}[]
\includegraphics[width=8.5cm]{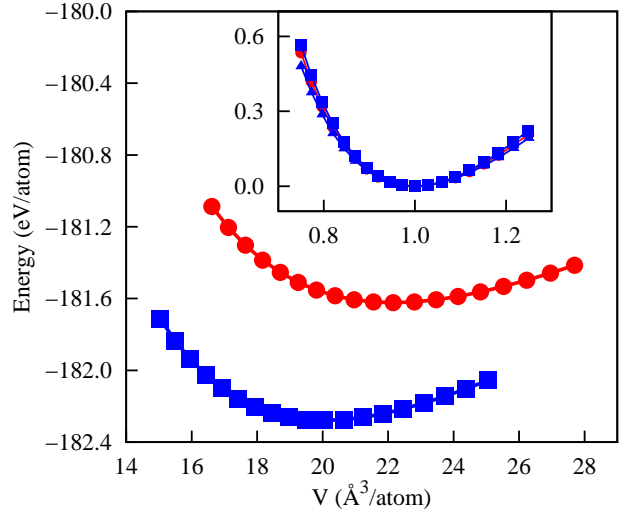}
%\scalebox{1.4}{\includegraphics{GaAs_eos.eps}}
\caption{\footnotesize (Color figures available online.) The equation of states (EOS) of ZB GaAs, using KS-DFT with LDA exchange functional and three different pseudopotentials, including Trouliers-Martins pseudopotential TM-NCPP (square), OEPP (circle) and BLPS (triangle). The inset show the EOS of total energies shifted by the equilibrium total energy as functions of the atomic volumes scaled by the equilibrium atomic volume. \label{GaAs_eos}}
\end{figure}

Next, we test the bulk properties of selected elemental solids and binary compounds. For each material, we first optimize its geometry at 0 GPa. After that, the volumes of the cells are changed from 0.95 $V_{0}$ to 1.05 $V_{0}$, and the calculated total energies at each volume is fitted by Birch-Murnaghan 3rd order EOS\cite{3B} to yield the bulk modulus. The energies as function of volumes are shown in Figs. \ref{Mg_eos} and \ref{GaAs_eos} for Mg in hexagonal-close-packing (HCP) structure and GaAs in Zinc Blende (ZB) structure. The calculated cohesive energies for many elemental solids or the enthalpy of formations for a number of binary compounds are shown in Tables \ref{EC} and \ref{CS}, together with the equilibrium volume and the bulk moduli and their pressure derivatives. 

\begin {table*}
\begin{ruledtabular}
\caption{ KS-DFT+OEPP predictions of the bulk properties for a number of elements, including the equilibrium volume ($V_{0}$), the bulk modulus ($B_{0}$), the  equilibrium total energy ($E_{0}$) and the cohesive energy $E_{c}$.  LDA exchange-correlation function is used. \label{EC} }  
\begin{tabular}{cccrrrrrr}

System&Structures &PPs&$E_{0}(eV)$&$E_{c}(eV)$&$V_{0}(\AA^{3})$&$V_{exp}(\AA^{3})\footnotemark[1]$ & $B_{0}(GPa)$&$B_{0}^{'}$\\
\hline
Li &bcc  &NCPP &-8.5265 &-2.0599 &18.5793 & &15.0 &3.2992\\
\multicolumn{2}{c}{} &OEPP &-9.2786 &-1.5259 &18.0179 & &16.5 &3.5459\\
\\
Na &bcc &NCPP &-102.2505 &-1.4174 &33.3281&39.4933&9.1 &3.3497\\
\multicolumn{2}{c}{} &OEPP &-102.2933 &-1.4479 &32.4215 & &9.5 &0.6122\\ 
\\
K  &bcc &NCPP &-18.0151 &-1.0333 &64.8243&75.2843 &4.6 &3.8632\\
\multicolumn{2}{c}{} &OEPP &-18.0093 & -1.1134 &63.0346 & &4.9 &3.9196\\ 
\\
Mg &hcp &NCPP &-24.6368 &-1.8192 &20.8056&23.2400 &39.7&3.8592\\
\multicolumn{2}{c}{} &NCPP(PBE) & &-1.5200 &22.6020 & &36.2 &\\
\multicolumn{2}{c}{} &OEPP &-24.5898 &-1.5455 &22.0320 & &36.6 &4.1335\\ 
\multicolumn{2}{c}{}&BLPS&&&21.175&&38\\
\\
Ca &fcc &NCPP &-39.0517 &-2.0845 &39.6116 &43.4819 &20.1 &3.4662\\
\multicolumn{2}{c}{} &OEPP &-38.5468 &-1.8520 &45.8677 & &19.8 &4.1649\\
\\
Al &fcc &NCPP &-57.2070 &-4.2334 &15.5407&16.6013 &84.1 &4.1211\\
\multicolumn{2}{c}{} &OEPP &-56.7969 &-3.6493 &18.3104 & &58.4 &5.1277\\ 
\multicolumn{2}{c}{}&BLPS &&&15.623&&84\\
\\
Si &CD &NCPP & -108.1010 &-6.0072 &19.4620 &20.0210 &96.3 &4.1806\\
\multicolumn{2}{c}{} &OEPP &-107.7690 &-5.0208 &22.9126 & &61.7 &4.2380\\ 
\\
P  &orthorhombic &NCPP &-180.4319 &-5.9977 &19.0757&19.0280 &81.0 &4.4198\\
\multicolumn{2}{c}{} &OEPP &-181.3171 &-5.3429 &23.3012 & &61.5 &3.8757\\
\multicolumn{2}{c}{}&BLPS&&&13.957&&133\\
\\
Ga &orthorhombic &NCPP &-106.9611 &-3.6216 &18.2695 &19.4690 &66.6 &5.7115\\
\multicolumn{2}{c}{} &OEPP &-107.4471 &-4.0423 &16.1598 & &83.4 &4.7387\\ 
\multicolumn{2}{c}{} &BLPS&&&17.232&&60\\
\\
Ge &CD &NCPP &-108.8499 &-4.9892 &22.9649 &22.6350 &65.0 &4.8198\\
\multicolumn{2}{c}{} &OEPP &-110.0396 &-5.3931 &20.8280 & &69.7 &-20.0241\\
\\
As &trigonal &NCPP &-255.5401 &-5.2602 &21.3964  &21.5210 &80.8 &4.1800\\
\multicolumn{2}{c}{} &OEPP &-256.3345 &-5.7947 &18.8517 & &91.6 &4.2379\\ 
\multicolumn{2}{c}{} &OEPP(PBE) &-255.7398 &-4.9241 &20.3134 & &76.0 &4.1581\\
\multicolumn{2}{c}{}&BLPS&&&20.033&&77\\
\\
Se &trigonal &NCPP &-258.6114 &-4.0206 &25.8296&27.2610 &63.0&4.2795\\
\multicolumn{2}{c}{} &OEPP &-259.5605 &-4.3883 &23.6493 & &69.8 &4.1601\\ 
\\
Br &orthorhombic &NCPP &-367.0428 &-2.0395 &32.8105 & &32.6 &4.5644\\
\multicolumn{2}{c}{} &OEPP &-367.9349 &-2.1755 &31.5890 & &34.7 &4.4997\\ 
\\
In &tetragonal &NCPP &-209.6031 &-3.3316 &22.6398 &26.1585 &56.9&5.2405\\
\multicolumn{2}{c}{} &OEPP &-210.7787 &-4.3875 &16.6604 & &94.9 &5.6000\\ 
\multicolumn{2}{c}{}&BLPS&&&20.052&&64\\
\\
Sb & trigonal &NCPP&-153.6718 &-4.9875 &28.4137 &30.2060 &63.6 &4.4383\\
\multicolumn{2}{c}{} &OEPP &-154.0335 &-5.9311 &23.0326 & &81.8 &3.1456\\
\multicolumn{2}{c}{}&BLPS&&&26.816&&63\\
\\
Te &trigonal &NCPP&-224.5969 &-3.8268 &32.0412&33.9250 &56.0 &4.5338\\
\multicolumn{2}{c}{} &OEPP &-225.0817 &-4.3429 &26.2784 & &67.4 &3.0025\\ 
\\

Zn &hcp &NCPP &-230.8038 &-2.0237 &12.4261 &15.212 &96.5 &4.7049\\
\multicolumn{2}{c}{} &OEPP &-231.8834 &-2.7594 &10.1611 & &129.1 &4.5924\\
\end{tabular}
\end{ruledtabular}
\footnotetext[1]{From Ref.~\cite{ex}}
\end{table*}

As shown in Fig. \ref{Mg_eos} and Fig. \ref{GaAs_eos}, both OEPP equilibrium volume and its dependence on the pressure deviates from the NCPP results. Similar to band structure, the OEPP EOS for Mg is close to that of BLPS. However, while the volume is scaled by the equilibrium volume ($V/V_{0}$), and the energy is aligned at the minimum point ($E-E_{0}$), the scaled EOS calculated from NCPP, OEPP and BLPS are very close and almost overlap, for both Mg and GaAs. 

As shown in Table \ref{EC}, the OEPP cohesive energies of s block elements, such as Li, Mg, Na etc, are generally lower than the NCPP results. In contrast, for p block elements, the OEPP results are higher in comparison with the NCPP results. There is no clear trend for the equilibrium volume, but the OEPP results are generally lower than the NCPP results, and they compares better for s block elements than p block elements. The available BLPS equilibrium volumes are closer to the NCPP volumes than OEPP; however, the BLPS potentials are constructed by fitting the bulk properties. Similar trend can be found for bulk modulus. 

\begin {table*}
\begin{ruledtabular}
\caption{KS-DFT+OEPP predictions of the bulk properties for selected binary compounds, including the equilibrium volume ($V_{0}$), the bulk modulus ($B_{0}$), the  equilibrium total energy ($E_{0}$) and the formation energy $E_{f}$. LDA is used for the exchange correlation functional.
\label{CS}. All the BLPS results ref. to \cite{Huang2010}}.  
\begin{tabular}{cccrrrrrr}

System&Structures &PPs&$E_{0}(eV)$&$E_{c}(eV)$&$V_{0}(\AA^{3})$&$V_{exp}(\AA^{3})\footnotemark[1]$&$B_{0}(GPa)$&$B_{0}^{'}$\\
\hline
GaP  &CD &NCPP  &-288.314 &-1.519 &39.228 &40.481 &91.5 &4.468\\
\multicolumn{2}{c}{} &OEPP  &-289.427 &-1.442 &39.956 & &79.2 &4.472\\
\multicolumn{2}{c}{} &BLPS  && &37.646 & &80 &\\
\\
GaAs &CD &NCPP &  -363.246 &-0.744 &44.310 &45.166 &73.5&4.483\\
\multicolumn{2}{c}{} &OEPP &-364.558 &-0.776 &40.105 & &86.7 &4.590\\
\multicolumn{2}{c}{} &BLPS  & &&40.634& &75 &4.472\\
\\
GaSb &CD &NCPP&-260.885 &-0.252 &54.523&56.617 &57.5&4.592\\
\multicolumn{2}{c}{} &OEPP  &-261.496 &-0.157 &48.918 & &71.3&4.576\\
\multicolumn{2}{c}{} &BLPS && &52.488 & &56 &\\
\\
MgSe &fcc &NCPP  &-285.941 &-2.693 &38.611  &40.492 &66.8 &3.998\\
\multicolumn{2}{c}{} &OEPP &-286.394 &-2.244 &40.089 & &66.6&4.156\\
\\
MgTe &hcp &NCPP  &-252.944 &-3.710 &62.880 &64.846 &38.8&4.086\\
\multicolumn{2}{c}{} &OEPP  &-251.879 &-2.208 &62.068 & &41.9 &4.134\\
\\
ZnSe &CD &NCPP& -491.150 &-1.735 &42.077&45.513 &76.2 &\\
\multicolumn{2}{c}{} &OEPP  &-493.377 &-1.933 &35.270 & &99.7 &4.338\\
\\
ZnTe &CD &NCPP  &-456.335 &-0.934 &52.806&56.773 &57.6 &4.478\\
\multicolumn{2}{c}{} &OEPP &-458.554 &-1.589 &43.940 & &80.2&4.772\\
\\
AlAs &CD &NCPP  &-313.728 &-0.981 &44.715 &45.383 &73.8 &4.162\\
\multicolumn{2}{c}{} &OEPP &-313.815 &-0.684 &46.128 & &70.5&4.297\\
\multicolumn{2}{c}{}&BLPS&&&43.616&&80
\end{tabular}
\end{ruledtabular}
\footnotetext[1]{From Ref~\cite{ex}}
\end{table*}

 \begin{table*}[!h]
\begin{ruledtabular}
\caption{Comparing OF-DFT and KS-DFT results of bulk moduli ($B_{0}$ in GPa), bulk equilibrium volumes ($V_{0}$ in $\AA^{3}$ ) and equilibrium total energies ($E_{0}$ in eV per atom) for Mg and Al. A number of structure are selected including hcp, simple cubit (sc), body-centered cubic (bcc) and fcc. The energies are refer to the energies of the equilibrium ground state structures, which are hcp for Mg and ccc for Al. Results obtained by five method combinations are compared, including KS-DFT+NCPP+LDA(or PBE ), KS-DFT+OEPP+LDA(or PBE) and OF-DFT+OEPP+LDA (using WGC KEDF).  \label{mg_al}}.
\begin{tabular}{clllll}
Mg && hcp & sc & bcc & fcc  \\ \hline  

  $V_{0}$ & KS-NCPP (PBE) & 20.835 (22.602)  &24.968 (27.257)  & 20.748 (22.560) & 20.970 (22.777)  \\
                 &  KS-OEPP (PBE) & 22.023 (23.606) &26.391 (28.478)  & 22.002 (23.656)  & 22.210 (23.818)  \\
                 &  OF-OEPP  & 22.225                &26.726             & 22.226               & 22.333   \\
                             \\
                             
  $a_{0}$ & KS-NCPP (PBE) & 3.093 (3.188)  &2.922 (3.010)   & 3.462 (3.561)  & 4.377 (4.499)  \\
  $c_{0}$ &  &5.027 (5.134)&&&\\
                 & KS-OEPP (PBE) &3.149 (3.237)   & 2.977 (3.054)  & 3.530 (3.618)  & 4.462 (4.569)  \\
                &      &5.126 (5.206)&&&\\
                & OF-OEPP   &3.154                &2.989            & 3.542               & 4.470\\
                & &5.158&&&\\ 
                \\
                
$E_{0}$ & KS-NCPP  (PBE) &0.000 (0.000) &  0.405 (0.389)  & 0.026 (0.029)  &0.010 (0.013)  \\
                 & KS-OEPP  (PBE) &0.000 (0.000) & 0.337 (0.334)   & 0.028 (0.030) &0.007 (0.010)  \\
                 &  OF-OEPP  &0.000    &0.322&0.024  &0.006   \\
                 \\
                 
$B_{0}$ & KS-NCPP (PBE) &41.4 (36.2)   &27.2 (22.6) & 43.6 (35.4) & 38.9 (35.2)  \\
               & KS-OEPP  (PBE) &36.5 (33.6) & 22.4 (21.2)  & 39.3 (32.7) & 35.9 (32.9)   \\
               & OF-OEPP   & 35.0           &    22.6      &  34.4          &  34.6\\ 
               \\ \hline
            
Al&&  fcc  &  bcc  &   sc &  hcp  \\ \hline

  $V_{0}$ &  KS-NCPP  (PBE)&    15.544 (16.525)&  15.906 (16.945) &19.055 (20.149) & 15.665 (17.922)\  \\
                 &   KS-OEPP (PBE)  &  18.029 (19.163) & 18.770 (19.615) & 21.218 (22.324)  & 18.488 (20.328)  \\
                 &   OF-OEPP  &  18.435               &  18.723               & 21.683&18.513 \\        
                     \\
  $a_{0}$  & KS-NCPP (PBE)  & 3.961 (4.044)  & 3.168 (3.238)  &  2.671 (2.722) & 2.798 (2.858)  \\
  $c_{0}$  &  &&&&4.618 (4.723)\\
                 & KS-OEPP (PBE)  &  4.162 (4.249)   & 3.348 (3.399)  &2.768 (2.817) & 2.959 (3.009)  \\
          &      &&&&4.876 (4.927)\\
                 &OF-OEPP   &4.192                & 3.345              &2.788 &2.960\\
          &    &&&&4.878\\
       \\
 $E_{0}$ &   KS-NCPP (PBE)  & 0.000 (0.000)  & 0.106 (0.094)  &  0.401 (0.371) &0.037 (0.031)  \\
                 &   KS-OEPP  (PBE) & 0.000 (0.000)  &0.050 (0.048)  &0.254 (0.246)  &0.014 (0.014)  \\
                 &   OF-OEPP   &0.000&0.054&0.223&0.011\\           
                  \\
 $B_{0}$ &   KS-NCPP (PBE) &  83.1 (77.2)  &  76.2 (67.3)  &   61.3 (55.8)&   79.3 (67.6) \\
                &   KS-OEPP  (PBE) &   69.4 (54.3)  &  54.9 (52.3) & 52.2 (45.8)  &   57.6 (50.8)   \\
                &   OF-OEPP   &    67.4          &   54.8           &50.6  &  56.3\\ 
                             \\   
\end{tabular}
\end{ruledtabular}
\end{table*}

For binary compounds, as shown in Table \ref{CS}, the OEPP results generally compare better with the NCPP results than for elemental solids. Furthermore, the available BLPS results are often worse than OEPP. This might be caused by the fact that the BLPS potentials are constructed for the elemental solids of which the chemical environment is very different. On the other hand, the construction of the OEPP can maximize the transferability. 

In order to test the OEPP in the OF-DFT scheme, we also performed OFDFT calculations for Mg and Al crystals. The bulk properties of Mg and Al in four different structures, including simple cubic (SC), body-centered-cubic (BCC), face-centered-cubic (FCC) and hexagonal-closed-packed (HCP) structures are calculated using both LDA and PBE-GGA exchange correlation functionals and presented in Table \ref{mg_al}.  As shown in the table, the difference between KS and OF-DFT is less significant than that between the OEPP and NCPP. However, only local pseudopotentials such as OEPP can be used in the OF-DFT calculations. Although the absolute values of OEPP show considerable discrepancy comparing with NCPP results, both KS-DFT and OF-DFT using OEPP reproduce the correct order for equilibrium volume, cohesive energy and bulk modulus of Mg and Al in different structures. The errors caused by use of local pseudopotentials are systematic. This indicates that OF-DFT using OEPP can be used for large scale simulation involving large number of structural configurations. 

\section{conclusion}

In summary, we proposed a systematic scheme of constructing local pseudopotentials directly from the electronic structure of atoms. This scheme is based on the optimized effective potential method and is found to be successful in generating local pseudopotnetials for large number of elements, with the accuracy and transferability close to the first principles pseudopotentials. For most of the elements in s and p-block except the second row, the LPP can be constructed and the test on real materials show that they can yield properties comparable to the empirical local pseudopotentials that are subtracted from the density functional calculations of the solid.   

The test results for atoms and bulk materials  also show that once the OEPP works well for the elements, it can also work well for the compounds formed by these elements. For many other elements, OEPP may yield large errors. In principle, the bulk properties can be restored by adjusting the construction parameters or adding corrections to the OEPP. However, it usually leads to local pseudopotentials that only work well at the chemical condition that it is fitted. Our practice reveals that the existence of a valid local pseudopotential with high transferability is an intrinsic property of the element. 

\begin{acknowledgments}
Y. M., W. M and S. Z. acknowledge the funding supports from China 973 Program under Grant No. 2011CB808200, National Natural Science Foundation of China under Grants No. 11274136, No. 11025418, and No. 91022029, 2012 Changjiang Scholar of Ministry of Education, and Changjiang Scholar and Innovative Research Team in University (Grant No. IRT1132). M.S.M. is supported by the MRSEC program (NSF- DMR1121053) and the ConvEne-IGERT Program (NSF-DGE 0801627). Part of the calculations was performed in the high performance computing center of Jilin University
\end{acknowledgments}

\appendix
\section{ Norm-conserving condition for local pseudopotentials}
Denoting the charge densities of each orbital for local and semi-local NC pseudopotentials as $n_{i}^{%
\text{LPP}}$ and $n_{i}^{\text{NCPP}}$, the norm-conserving condition is

\begin{equation}
\int_{\Omega }n_{i}^{\text{LPP}}\left( r\right) d^{3}r=\int_{\Omega }n_{i}^{%
\text{NCPP}}\left( r\right) d^{3}r.
\end{equation}

It iw worth to notice the analogy between the NC condition and the OEP method.
For OEP, the total density is conserved at each point in the real space
while in NC condition the integration of the density for each orbital inside
the core sphere is conserved. In both cases, the shifts of the orbitals are
first order. Denoting the orbitals of local and semi-local NC pseudopotentials as $\varphi _{i}\left(
r\right) $ and $\psi _{i}\left( r\right) $, and their differences  as 
$\tilde{\psi}_{i}\left( r\right) $, the NC condition can be rewritten
as:
\begin{equation}
\int_{\Omega }\tilde{\psi}_{i}^{\ast }\left( r\right) \varphi _{i
}\left( r\right)d^{3}r +\int_{\Omega }\varphi _{i}^{\ast }\left( r\right) 
\tilde{\psi}_{i}\left( r\right)d^{3}r =0.
\end{equation}

Assuming that the changes from semi-local pseudopotential to the local pseudopotential is 
a perturbation, the changes of the wavefunction $\tilde{\psi}_{i}\left( r\right) $ 
can be expressed through first order perturbation as:
\begin{equation}
\tilde{\psi}_{i}^{\ast }\left( r\right) =\sum_{j\neq i}\int 
\frac{\varphi _{j}^{\ast }\left( r^{\prime }\right) \left( v\left( r^{\prime
}\right) -v_{i}\left( r^{\prime }\right) \right) \varphi _{i}\left(
r^{\prime }\right) }{\varepsilon _{j}-\varepsilon _{i}}\varphi _{j}\left(
r\right) d^{3}r^{\prime}.
\end{equation}
Rewriting NC condition,
\begin{equation}
\int_{\Omega }\tilde{\psi}_{i}^{\ast }\left( r\right) \varphi _{i
}\left( r\right)d^{3}r +c.c.=0
\end{equation}
as 
\begin{equation}
\int_{\Omega }\tilde{\psi}_{i}^{\ast }\left( r\right) \varepsilon
_{i}\varphi _{i}\left( r\right)d^{3}r +c.c.=0
\end{equation}, 
which can be further transformed to
\begin{equation}
\int_{\Omega }\tilde{\psi}_{i}^{\ast }\left( r\right) \left( -\frac{1}{2}%
\nabla ^{2}+v_{i}\left( r\right) \right) \varphi _{i}\left( r\right)d^{3}r
+c.c. = 0
\end{equation}
 by employing KS equation
\begin{equation}
\left( -\frac{1}{2}\nabla ^{2}+v_{i}\left( r\right) \right) \varphi _{i}\left( r\right)
=\varepsilon _{i}\varphi _{i}(r). 
\end{equation} 
Using partial integration, we have 
\begin{widetext}
\begin{eqnarray}
\int_{\Omega }\tilde{\psi}_{i}^{\ast }\left( r\right) \nabla ^{2}\varphi
_{i}\left( r\right)d^{3}r  &=&\tilde{\psi}_{i}^{\ast }\left( r\right)
\nabla \varphi _{i}\left( r\right) |_{\Omega }-\int_{\Omega }\nabla 
\tilde{\psi}_{i}^{\ast }\left( r\right) \nabla \varphi _{i}\left(
r\right)d^{3}r  \\ \nonumber
&=&\tilde{\psi}_{i}^{\ast }\left( r\right) \nabla \varphi _{i}\left(
r\right) |_{\Omega }-\nabla \tilde{\psi}_{i}^{\ast }\left( r\right) \varphi
_{i}\left( r\right) |_{\Omega }+\int_{\Omega }\nabla ^{2}\tilde{\psi%
}_{i}^{\ast }\left( r\right) \varphi _{i}\left( r\right)d^{3}r .
\end{eqnarray}
The first two terms should be exactly 0 while the NC condition is satisfied 
because the local pseudopotential will now produces wavefunctions that identical to the 
wavefunctions of all electron potential as well as the NC pseudopotentials outside the core
region. 
Therefore, $\tilde{\psi}_{i}^{\ast }\left(
r\right) $ and $\nabla \tilde{\psi}_{i}^{\ast }\left( r\right) $ should be 0
at the core sphere and in the region out of the core. 

As proved by KLI (see Eqn. 59 in Ref. \cite{KLI1995}), 
\begin{equation}
-\frac{1}{2}\nabla ^{2}\tilde{\psi}_{i}^{\ast }\left( r\right) =\left[
v\left( r\right) -v_{i}\left( r\right) -\left( \bar{v}_{i}-\bar{v}%
_{i}^{i}\right) \right] \varphi _{i}^{\ast }\left( r\right) +\left(
\varepsilon _{i}-v_{i}\left( r\right) \right) \tilde{\psi}_{i}^{\ast }\left(
r\right) .
\end{equation}
Thus, we can transform the NC condition as  
\begin{eqnarray}
\int_{\Omega }\tilde{\psi}_{i}^{\ast }\left( r\right) \left( -\frac{1}{2}
\nabla ^{2}+v_{i}\left( r\right) \right) \varphi _{i}\left( r\right)d^{3}r
&=&\int_{\Omega }\left\{ \left( -\frac{1}{2}\nabla ^{2}\tilde{\psi}
_{i}^{\ast }\left( r\right) \right) \varphi _{i}\left( r\right) +\tilde{\psi}
_{i}^{\ast }\left( r\right) v_{i}\left( r\right) \varphi _{i}\left( r\right)
\right\}d^{3}r \nonumber  \\   \nonumber
&=&\int_{\Omega }\{\left[ v\left( r\right) -v_{i}\left( r\right) -\left( 
\bar{v}_{i}-\bar{v}_{i}^{i}\right) \right] \tilde{\varphi}_{i}^{\ast }\left(
r\right) \varphi _{i}\left( r\right) \\ \nonumber
&&+\left( \varepsilon _{i}-v_{i}\left(
r\right) \right) \tilde{\psi}_{i}^{\ast }\left( r\right) \varphi
_{i}\left( r\right)  
+\tilde{\psi}_{i}^{\ast }\left( r\right) v_{i}\left( r\right) \varphi
_{i}\left( r\right) \}d^{3}r \\  \nonumber
&=&\int_{\Omega }\{\left[ v\left( r\right) -v_{i}\left( r\right) -\left( 
\bar{v}_{i}-\bar{v}_{i}^{i}\right) \right] \tilde{\varphi}_{i}^{\ast }\left(
r\right) \varphi _{i}\left( r\right)d^{3}r  \\   \nonumber
&=&0
\end{eqnarray}%\end{widetext}
%%&&&&&&&&&&&&&&&&&&&&&&&&&
This means that 
\begin{equation}
\int_{\Omega }n_{i}\left( r\right) v\left( r\right) dr-\int_{\Omega
}n_{i}\left( r\right) v_{i}\left( r\right) dr-\int_{\Omega }n_{i}\left(
r\right) \left( \bar{v}^{i}-\bar{v}_{i}^{i}\right) dr = 0
\end{equation}%%%%%%%%
Recalling that $\bar{v}_{i}^{i}=\int_{\Omega }n_{i}\left( r\right)
v_{i}\left( r\right) dr$ and $\bar{v}^{i}=\int_{\Omega }n_{i}\left( r\right)
v\left( r\right) dr$, the above equation can be written as 
\begin{equation}
\left( \bar{v}^{i}-\bar{v}_{i}^{i}\right) =\left( \bar{v}^{i}-\bar{v}%
_{i}^{i}\right) \int_{\Omega }n_{i}\left( r\right) dr=\left( \bar{v}^{i}-%
\bar{v}_{i}^{i}\right) \bar{n}_{i}
\end{equation}
Because $\bar{n}_{i}$ is the partial charge enclosed in the core region and should be 
smaller than 1, $\left( \bar{v}^{i}-\bar{v}_{i}^{i}\right) =0,$which can be more
clearly shown as: 
\begin{equation}
\int_{\Omega }n_{i}\left( r\right) \left( v\left( r\right) -v_{i}\left(
r\right) \right) dr=0.
\end{equation}
Thus, we prove Eqn.\ref{nv} in Section II.
\end{widetext}
\begin{widetext}
\section{The  electronic structure of (pseudo) atom and related definitions \label{appB}}
For a given set of valence occupancies $f_{l}$ the pseudo valence states are determined by self-consistently solving the radial Schr\"dingier equations associated with the pseudopotential
\begin{equation}
\left[ -\frac{1}{2}\frac{d^2}{dr^2} + \frac{l(l+1)}{2r^2} 
+ V^{\rm HXC}(r) + V^{\rm ps}_{l}(r)
- \epsilon^{\rm ps}_l \right] u^{\rm ps}_{l}(r) = 0 \quad,
\end{equation}  
with the screening potential
\begin{equation}
V^{\rm HXC}(r)
= V^{\rm XC}[\rho^{\rm ps}+\tilde\rho^{\rm core}_0;r] + V^{\rm H}[\rho^{\rm ps};r]
\quad,
\end{equation}
and 
$\rho^{\rm ps}(r) = \frac{1}{4\pi r^2}\sum_{l} f_{l} \left|u^{\rm ps}_{l}(r)\right|^2$. The total energy of the pseudo atom is given by
%
%\begin{widetext}
\begin{equation}
E^{\rm tot-PS}[\rho^{\rm ps}] = 
T[\rho^{\rm ps}] + E^{\rm XC}[\rho^{\rm ps}+\tilde\rho^{\rm core}_{0}] + E^{\rm H}[\rho^{\rm ps}] + 
\sum_{l} f_{l} \int_{0}^{\infty} V^{\rm ps}_{l}(r) |u^{\rm ps}_{l}(r)|^2\, dr \quad,
\end{equation}
where the kinetic energy associated with the pseudo valence states is
\begin{equation}
T[\rho^{\rm ps}] = \sum_{l} f_{l} \left( \epsilon^{\rm ps}_l 
- \int_{0}^{\infty} \left\{ V^{\rm HXC}(r)
+ V^{\rm ps}_{l}(r) \right\} |u^{\rm ps}_{l}(r)|^2\, dr \right) .
\end{equation}
The above equations refer to Eq.(51-54) in Ref. \cite{FHI98PP} 
Here several density related quantities are defined as follows:
\begin{equation}
\rho^{c}_{l}=\int_{0}^{R_{cut}}\ |u^{ps}_{l}(\epsilon_{l}^{ps};r)\ |^{2}dr
\end{equation}
It has the following relation for non-local normal-conserving pseudopotential.  
\begin{equation}
\rho^{c}_{l}=\int_{0}^{R_{cut}} |u^{ps}_{l}(\epsilon_{l}^{ps};r) |^{2}dr = \int_{0}^{R_{cut}} |u^{ae}_{l}(\epsilon_{l}^{ae};r) |^{2}dr
\end{equation}
Finally, we define $\delta_{\rho}$
\begin{equation}
  \delta_{\rho}=\int_{\Omega}\ |n_{\text{PS}}(r)-n_{\text{TM}}(r)\ |d^{3}r 
\end{equation}
\end{widetext}

\nocite{*}

\bibliography{OEPP}% Produces the bibliography via BibTeX.

\begin{thebibliography}{44}
\expandafter\ifx\csname natexlab\endcsname\relax\def\natexlab#1{#1}\fi
\expandafter\ifx\csname bibnamefont\endcsname\relax
  \def\bibnamefont#1{#1}\fi
\expandafter\ifx\csname bibfnamefont\endcsname\relax
  \def\bibfnamefont#1{#1}\fi
\expandafter\ifx\csname citenamefont\endcsname\relax
  \def\citenamefont#1{#1}\fi
\expandafter\ifx\csname url\endcsname\relax
  \def\url#1{\texttt{#1}}\fi
\expandafter\ifx\csname urlprefix\endcsname\relax\def\urlprefix{URL }\fi
\providecommand{\bibinfo}[2]{#2}
\providecommand{\eprint}[2][]{\url{#2}}

\bibitem[{\citenamefont{Wang and Carter}(2002)}]{WangCarter2000}
\bibinfo{author}{\bibfnamefont{Y.~A.} \bibnamefont{Wang}} \bibnamefont{and}
  \bibinfo{author}{\bibfnamefont{E.~A.} \bibnamefont{Carter}}, in
  \emph{\bibinfo{booktitle}{Theoretical methods in condensed phase chemistry}}
  (\bibinfo{publisher}{Springer}, \bibinfo{year}{2002}), pp.
  \bibinfo{pages}{117--184}.

\bibitem[{\citenamefont{Wesolowski}(2013)}]{wesolowski2013}
\bibinfo{author}{\bibfnamefont{T.~A.} \bibnamefont{Wesolowski}},
  \emph{\bibinfo{title}{Recent progress in orbital-free density functional
  theory (recent advances in computational chemistry)}}
  (\bibinfo{publisher}{World Scientific Publishing Company},
  \bibinfo{year}{2013}).

\bibitem[{\citenamefont{Shin and Carter}(2012)}]{Shin2012}
\bibinfo{author}{\bibfnamefont{I.}~\bibnamefont{Shin}} \bibnamefont{and}
  \bibinfo{author}{\bibfnamefont{E.~A.} \bibnamefont{Carter}},
  \bibinfo{journal}{Modelling and Simulation in Materials Science and
  Engineering} \textbf{\bibinfo{volume}{20}}, \bibinfo{pages}{015006}
  (\bibinfo{year}{2012}).

\bibitem[{\citenamefont{Vora}(2010)}]{Vora2010}
\bibinfo{author}{\bibfnamefont{A.~M.} \bibnamefont{Vora}},
  \bibinfo{journal}{Physics and Chemistry of Liquids}
  \textbf{\bibinfo{volume}{48}}, \bibinfo{pages}{723} (\bibinfo{year}{2010}).

\bibitem[{\citenamefont{Lign脙拧res and Carter}(2005)}]{Ligneres2007}
\bibinfo{author}{\bibfnamefont{V.}~\bibnamefont{Lign脙拧res}}
  \bibnamefont{and} \bibinfo{author}{\bibfnamefont{E.}~\bibnamefont{Carter}},
  in \emph{\bibinfo{booktitle}{Handbook of Materials Modeling}}, edited by
  \bibinfo{editor}{\bibfnamefont{S.}~\bibnamefont{Yip}}
  (\bibinfo{publisher}{Springer Netherlands}, \bibinfo{year}{2005}), pp.
  \bibinfo{pages}{137--148}, ISBN \bibinfo{isbn}{978-1-4020-3287-5}.

\bibitem[{\citenamefont{Zhou et~al.}(2005)\citenamefont{Zhou, Ligneres, and
  Carter}}]{Zhouj2005}
\bibinfo{author}{\bibfnamefont{B.}~\bibnamefont{Zhou}},
  \bibinfo{author}{\bibfnamefont{V.~L.} \bibnamefont{Ligneres}},
  \bibnamefont{and} \bibinfo{author}{\bibfnamefont{E.~A.}
  \bibnamefont{Carter}}, \bibinfo{journal}{The Journal of Chemical Physics}
  \textbf{\bibinfo{volume}{122}}, \bibinfo{eid}{044103} (\bibinfo{year}{2005}).

\bibitem[{\citenamefont{Bhatt et~al.}(2005)\citenamefont{Bhatt, Vyas, Jani, and
  Gohel}}]{Bhatt2005}
\bibinfo{author}{\bibfnamefont{N.}~\bibnamefont{Bhatt}},
  \bibinfo{author}{\bibfnamefont{P.}~\bibnamefont{Vyas}},
  \bibinfo{author}{\bibfnamefont{A.}~\bibnamefont{Jani}}, \bibnamefont{and}
  \bibinfo{author}{\bibfnamefont{V.}~\bibnamefont{Gohel}},
  \bibinfo{journal}{Journal of Physics and Chemistry of Solids}
  \textbf{\bibinfo{volume}{66}}, \bibinfo{pages}{797 } (\bibinfo{year}{2005}),
  ISSN \bibinfo{issn}{0022-3697}.

\bibitem[{\citenamefont{Jiang and Yang}(2004)}]{Jiang2004}
\bibinfo{author}{\bibfnamefont{H.}~\bibnamefont{Jiang}} \bibnamefont{and}
  \bibinfo{author}{\bibfnamefont{W.}~\bibnamefont{Yang}}, \bibinfo{journal}{The
  Journal of Chemical Physics} \textbf{\bibinfo{volume}{121}},
  \bibinfo{pages}{2030} (\bibinfo{year}{2004}).

\bibitem[{\citenamefont{Gonz{\'a}lez et~al.}(2001)\citenamefont{Gonz{\'a}lez,
  Gonz{\'a}lez, L{\'o}pez, and Stott}}]{Gonzalez2001}
\bibinfo{author}{\bibfnamefont{D.~J.} \bibnamefont{Gonz{\'a}lez}},
  \bibinfo{author}{\bibfnamefont{L.~E.} \bibnamefont{Gonz{\'a}lez}},
  \bibinfo{author}{\bibfnamefont{J.~M.} \bibnamefont{L{\'o}pez}},
  \bibnamefont{and} \bibinfo{author}{\bibfnamefont{M.~J.} \bibnamefont{Stott}},
  \bibinfo{journal}{The Journal of Chemical Physics}
  \textbf{\bibinfo{volume}{115}}, \bibinfo{pages}{2373} (\bibinfo{year}{2001}).

\bibitem[{\citenamefont{Pearson et~al.}(1993)\citenamefont{Pearson, Smargiassi,
  and Madden}}]{Pearson1993}
\bibinfo{author}{\bibfnamefont{M.}~\bibnamefont{Pearson}},
  \bibinfo{author}{\bibfnamefont{E.}~\bibnamefont{Smargiassi}},
  \bibnamefont{and} \bibinfo{author}{\bibfnamefont{P.}~\bibnamefont{Madden}},
  \bibinfo{journal}{Journal of Physics: Condensed Matter}
  \textbf{\bibinfo{volume}{5}}, \bibinfo{pages}{3221} (\bibinfo{year}{1993}).

\bibitem[{\citenamefont{Huang and Carter}(2010)}]{Huang2010}
\bibinfo{author}{\bibfnamefont{C.}~\bibnamefont{Huang}} \bibnamefont{and}
  \bibinfo{author}{\bibfnamefont{E.~A.} \bibnamefont{Carter}},
  \bibinfo{journal}{Phys. Rev. B} \textbf{\bibinfo{volume}{81}},
  \bibinfo{pages}{045206} (\bibinfo{year}{2010}).

\bibitem[{\citenamefont{Xia et~al.}(2012)\citenamefont{Xia, Huang, Shin, and
  Carter}}]{xia2012}
\bibinfo{author}{\bibfnamefont{J.}~\bibnamefont{Xia}},
  \bibinfo{author}{\bibfnamefont{C.}~\bibnamefont{Huang}},
  \bibinfo{author}{\bibfnamefont{I.}~\bibnamefont{Shin}}, \bibnamefont{and}
  \bibinfo{author}{\bibfnamefont{E.~A.} \bibnamefont{Carter}},
  \bibinfo{journal}{The Journal of Chemical Physics}
  \textbf{\bibinfo{volume}{136}}, \bibinfo{eid}{084102} (\bibinfo{year}{2012}).

\bibitem[{\citenamefont{Goodwin et~al.}(1990)\citenamefont{Goodwin, Needs, and
  Heine}}]{Goodwin1990}
\bibinfo{author}{\bibfnamefont{L.}~\bibnamefont{Goodwin}},
  \bibinfo{author}{\bibfnamefont{R.}~\bibnamefont{Needs}}, \bibnamefont{and}
  \bibinfo{author}{\bibfnamefont{V.}~\bibnamefont{Heine}},
  \bibinfo{journal}{Journal of Physics: Condensed Matter}
  \textbf{\bibinfo{volume}{2}}, \bibinfo{pages}{351} (\bibinfo{year}{1990}).

\bibitem[{\citenamefont{Fiolhais et~al.}(1995)\citenamefont{Fiolhais, Perdew,
  Armster, MacLaren, and Brajczewska}}]{Fiolhais1995}
\bibinfo{author}{\bibfnamefont{C.}~\bibnamefont{Fiolhais}},
  \bibinfo{author}{\bibfnamefont{J.~P.} \bibnamefont{Perdew}},
  \bibinfo{author}{\bibfnamefont{S.~Q.} \bibnamefont{Armster}},
  \bibinfo{author}{\bibfnamefont{J.~M.} \bibnamefont{MacLaren}},
  \bibnamefont{and}
  \bibinfo{author}{\bibfnamefont{M.}~\bibnamefont{Brajczewska}},
  \bibinfo{journal}{Phys. Rev. B} \textbf{\bibinfo{volume}{51}},
  \bibinfo{pages}{14001} (\bibinfo{year}{1995}).

\bibitem[{\citenamefont{Fiolhais et~al.}(1996)\citenamefont{Fiolhais, Perdew,
  Armster, MacLaren, and Brajczewska}}]{Fiolhais1996}
\bibinfo{author}{\bibfnamefont{C.}~\bibnamefont{Fiolhais}},
  \bibinfo{author}{\bibfnamefont{J.~P.} \bibnamefont{Perdew}},
  \bibinfo{author}{\bibfnamefont{S.~Q.} \bibnamefont{Armster}},
  \bibinfo{author}{\bibfnamefont{J.~M.} \bibnamefont{MacLaren}},
  \bibnamefont{and}
  \bibinfo{author}{\bibfnamefont{M.}~\bibnamefont{Brajczewska}},
  \bibinfo{journal}{Phys. Rev. B} \textbf{\bibinfo{volume}{53}},
  \bibinfo{pages}{13193} (\bibinfo{year}{1996}).

\bibitem[{\citenamefont{Watson et~al.}(1998)\citenamefont{Watson, Jesson,
  Carter, and Madden}}]{Watson1998}
\bibinfo{author}{\bibfnamefont{S.}~\bibnamefont{Watson}},
  \bibinfo{author}{\bibfnamefont{B.}~\bibnamefont{Jesson}},
  \bibinfo{author}{\bibfnamefont{E.}~\bibnamefont{Carter}}, \bibnamefont{and}
  \bibinfo{author}{\bibfnamefont{P.}~\bibnamefont{Madden}},
  \bibinfo{journal}{EPL (Europhysics Letters)} \textbf{\bibinfo{volume}{41}},
  \bibinfo{pages}{37} (\bibinfo{year}{1998}).

\bibitem[{\citenamefont{Jesson and Madden}(2000)}]{Jesson2000}
\bibinfo{author}{\bibfnamefont{B.~J.} \bibnamefont{Jesson}} \bibnamefont{and}
  \bibinfo{author}{\bibfnamefont{P.~A.} \bibnamefont{Madden}},
  \bibinfo{journal}{The Journal of Chemical Physics}
  \textbf{\bibinfo{volume}{113}}, \bibinfo{pages}{5924} (\bibinfo{year}{2000}).

\bibitem[{\citenamefont{Wang and Stott}(2003)}]{Wang2003}
\bibinfo{author}{\bibfnamefont{B.}~\bibnamefont{Wang}} \bibnamefont{and}
  \bibinfo{author}{\bibfnamefont{M.~J.} \bibnamefont{Stott}},
  \bibinfo{journal}{Phys. Rev. B} \textbf{\bibinfo{volume}{68}},
  \bibinfo{pages}{195102} (\bibinfo{year}{2003}).

\bibitem[{\citenamefont{Zhou et~al.}(2004)\citenamefont{Zhou, Alexander~Wang,
  and Carter}}]{Zhou2004}
\bibinfo{author}{\bibfnamefont{B.}~\bibnamefont{Zhou}},
  \bibinfo{author}{\bibfnamefont{Y.}~\bibnamefont{Alexander~Wang}},
  \bibnamefont{and} \bibinfo{author}{\bibfnamefont{E.~A.}
  \bibnamefont{Carter}}, \bibinfo{journal}{Phys. Rev. B}
  \textbf{\bibinfo{volume}{69}}, \bibinfo{pages}{125109}
  (\bibinfo{year}{2004}).

\bibitem[{\citenamefont{Zhou and Carter}(2005)}]{Zhou2005}
\bibinfo{author}{\bibfnamefont{B.}~\bibnamefont{Zhou}} \bibnamefont{and}
  \bibinfo{author}{\bibfnamefont{E.~A.} \bibnamefont{Carter}},
  \bibinfo{journal}{The Journal of Chemical Physics}
  \textbf{\bibinfo{volume}{122}}, \bibinfo{eid}{184108} (\bibinfo{year}{2005}).

\bibitem[{\citenamefont{Huang and Carter}(2008)}]{Huang2008}
\bibinfo{author}{\bibfnamefont{C.}~\bibnamefont{Huang}} \bibnamefont{and}
  \bibinfo{author}{\bibfnamefont{E.~A.} \bibnamefont{Carter}},
  \bibinfo{journal}{Physical Chemistry Chemical Physics}
  \textbf{\bibinfo{volume}{10}}, \bibinfo{pages}{7109} (\bibinfo{year}{2008}).

\bibitem[{\citenamefont{Hamann et~al.}(1979)\citenamefont{Hamann, Schl\"uter,
  and Chiang}}]{Hamann1979}
\bibinfo{author}{\bibfnamefont{D.~R.} \bibnamefont{Hamann}},
  \bibinfo{author}{\bibfnamefont{M.}~\bibnamefont{Schl\"uter}},
  \bibnamefont{and} \bibinfo{author}{\bibfnamefont{C.}~\bibnamefont{Chiang}},
  \bibinfo{journal}{Phys. Rev. Lett.} \textbf{\bibinfo{volume}{43}},
  \bibinfo{pages}{1494} (\bibinfo{year}{1979}).

\bibitem[{\citenamefont{Troullier and Martins}(1991)}]{Troullier1991}
\bibinfo{author}{\bibfnamefont{N.}~\bibnamefont{Troullier}} \bibnamefont{and}
  \bibinfo{author}{\bibfnamefont{J.~L.} \bibnamefont{Martins}},
  \bibinfo{journal}{Phys. Rev. B} \textbf{\bibinfo{volume}{43}},
  \bibinfo{pages}{1993} (\bibinfo{year}{1991}).

\bibitem[{\citenamefont{Slater}(1951)}]{Slater1951}
\bibinfo{author}{\bibfnamefont{J.~C.} \bibnamefont{Slater}},
  \bibinfo{journal}{Phys. Rev.} \textbf{\bibinfo{volume}{81}},
  \bibinfo{pages}{385} (\bibinfo{year}{1951}).

\bibitem[{\citenamefont{Krieger et~al.}(1995)\citenamefont{Krieger, Li, and
  Iafrate}}]{KLI1995}
\bibinfo{author}{\bibfnamefont{J.}~\bibnamefont{Krieger}},
  \bibinfo{author}{\bibfnamefont{Y.}~\bibnamefont{Li}}, \bibnamefont{and}
  \bibinfo{author}{\bibfnamefont{G.}~\bibnamefont{Iafrate}},
  \bibinfo{journal}{Plenum Press, New York} p. \bibinfo{pages}{p191}
  (\bibinfo{year}{1995}).

\bibitem[{\citenamefont{Grabo et~al.}(2000)\citenamefont{Grabo, Kreibich,
  Kurth, and Gross}}]{Grabo1998}
\bibinfo{author}{\bibfnamefont{T.}~\bibnamefont{Grabo}},
  \bibinfo{author}{\bibfnamefont{T.}~\bibnamefont{Kreibich}},
  \bibinfo{author}{\bibfnamefont{S.}~\bibnamefont{Kurth}}, \bibnamefont{and}
  \bibinfo{author}{\bibfnamefont{E.}~\bibnamefont{Gross}},
  \bibinfo{journal}{Strong Coulomb Correlations in Electronic Structure
  Calculations: Beyond Local Density Approximations} p. \bibinfo{pages}{203}
  (\bibinfo{year}{2000}).

\bibitem[{\citenamefont{Bylander and Kleinman}(1995)}]{Bylander1995}
\bibinfo{author}{\bibfnamefont{D.~M.} \bibnamefont{Bylander}} \bibnamefont{and}
  \bibinfo{author}{\bibfnamefont{L.}~\bibnamefont{Kleinman}},
  \bibinfo{journal}{Phys. Rev. B} \textbf{\bibinfo{volume}{52}},
  \bibinfo{pages}{14566} (\bibinfo{year}{1995}).

\bibitem[{\citenamefont{Miao}(2000)}]{Miao2000}
\bibinfo{author}{\bibfnamefont{M.}~\bibnamefont{Miao}},
  \bibinfo{journal}{Philosophical Magazine B} \textbf{\bibinfo{volume}{80}},
  \bibinfo{pages}{409} (\bibinfo{year}{2000}).

\bibitem[{\citenamefont{K\"ummel and Perdew}(2003)}]{Kummel2003}
\bibinfo{author}{\bibfnamefont{S.}~\bibnamefont{K\"ummel}} \bibnamefont{and}
  \bibinfo{author}{\bibfnamefont{J.~P.} \bibnamefont{Perdew}},
  \bibinfo{journal}{Phys. Rev. B} \textbf{\bibinfo{volume}{68}},
  \bibinfo{pages}{035103} (\bibinfo{year}{2003}).

\bibitem[{\citenamefont{Fuchs and Scheffler}(1999)}]{FHI98PP}
\bibinfo{author}{\bibfnamefont{M.}~\bibnamefont{Fuchs}} \bibnamefont{and}
  \bibinfo{author}{\bibfnamefont{M.}~\bibnamefont{Scheffler}},
  \bibinfo{journal}{Computer Physics Communications}
  \textbf{\bibinfo{volume}{119}}, \bibinfo{pages}{67} (\bibinfo{year}{1999}).

\bibitem[{\citenamefont{Louie et~al.}(1982)\citenamefont{Louie, Froyen, and
  Cohen}}]{nlcc}
\bibinfo{author}{\bibfnamefont{S.~G.} \bibnamefont{Louie}},
  \bibinfo{author}{\bibfnamefont{S.}~\bibnamefont{Froyen}}, \bibnamefont{and}
  \bibinfo{author}{\bibfnamefont{M.~L.} \bibnamefont{Cohen}},
  \bibinfo{journal}{Phys. Rev. B} \textbf{\bibinfo{volume}{26}},
  \bibinfo{pages}{1738} (\bibinfo{year}{1982}).

\bibitem[{\citenamefont{Kleinman and Bylander}(1982)}]{KB}
\bibinfo{author}{\bibfnamefont{L.}~\bibnamefont{Kleinman}} \bibnamefont{and}
  \bibinfo{author}{\bibfnamefont{D.~M.} \bibnamefont{Bylander}},
  \bibinfo{journal}{Phys. Rev. Lett.} \textbf{\bibinfo{volume}{48}},
  \bibinfo{pages}{1425} (\bibinfo{year}{1982}).

\bibitem[{\citenamefont{Segall et~al.}(2002)\citenamefont{Segall, Lindan,
  Probert, Pickard, Hasnip, Clark, and Payne}}]{CASTEP}
\bibinfo{author}{\bibfnamefont{M.}~\bibnamefont{Segall}},
  \bibinfo{author}{\bibfnamefont{P.~J.} \bibnamefont{Lindan}},
  \bibinfo{author}{\bibfnamefont{M.~a.} \bibnamefont{Probert}},
  \bibinfo{author}{\bibfnamefont{C.}~\bibnamefont{Pickard}},
  \bibinfo{author}{\bibfnamefont{P.}~\bibnamefont{Hasnip}},
  \bibinfo{author}{\bibfnamefont{S.}~\bibnamefont{Clark}}, \bibnamefont{and}
  \bibinfo{author}{\bibfnamefont{M.}~\bibnamefont{Payne}},
  \bibinfo{journal}{Journal of Physics: Condensed Matter}
  \textbf{\bibinfo{volume}{14}}, \bibinfo{pages}{2717} (\bibinfo{year}{2002}).

\bibitem[{\citenamefont{Ceperley and Alder}(1980)}]{ldaca}
\bibinfo{author}{\bibfnamefont{D.~M.} \bibnamefont{Ceperley}} \bibnamefont{and}
  \bibinfo{author}{\bibfnamefont{B.~J.} \bibnamefont{Alder}},
  \bibinfo{journal}{Phys. Rev. Lett.} \textbf{\bibinfo{volume}{45}},
  \bibinfo{pages}{566} (\bibinfo{year}{1980}).

\bibitem[{\citenamefont{Perdew et~al.}(1996)\citenamefont{Perdew, Burke, and
  Ernzerhof}}]{pbe}
\bibinfo{author}{\bibfnamefont{J.~P.} \bibnamefont{Perdew}},
  \bibinfo{author}{\bibfnamefont{K.}~\bibnamefont{Burke}}, \bibnamefont{and}
  \bibinfo{author}{\bibfnamefont{M.}~\bibnamefont{Ernzerhof}},
  \bibinfo{journal}{Phys. Rev. Lett.} \textbf{\bibinfo{volume}{77}},
  \bibinfo{pages}{3865} (\bibinfo{year}{1996}).

\bibitem[{ATL()}]{ATLAS}
\bibinfo{note}{To be published}.

\bibitem[{\citenamefont{Wang et~al.}(1999)\citenamefont{Wang, Govind, and
  Carter}}]{WGC99}
\bibinfo{author}{\bibfnamefont{Y.~A.} \bibnamefont{Wang}},
  \bibinfo{author}{\bibfnamefont{N.}~\bibnamefont{Govind}}, \bibnamefont{and}
  \bibinfo{author}{\bibfnamefont{E.~A.} \bibnamefont{Carter}},
  \bibinfo{journal}{Phys. Rev. B} \textbf{\bibinfo{volume}{60}},
  \bibinfo{pages}{16350} (\bibinfo{year}{1999}).

\bibitem[{\citenamefont{Nagy}(1997)}]{Nagy1997}
\bibinfo{author}{\bibfnamefont{A.}~\bibnamefont{Nagy}}, \bibinfo{journal}{Phys.
  Rev. A} \textbf{\bibinfo{volume}{55}}, \bibinfo{pages}{3465}
  (\bibinfo{year}{1997}).

\bibitem[{\citenamefont{Bachelet et~al.}(1982)\citenamefont{Bachelet, Hamann,
  and Schl\"uter}}]{Bachelet1982}
\bibinfo{author}{\bibfnamefont{G.~B.} \bibnamefont{Bachelet}},
  \bibinfo{author}{\bibfnamefont{D.~R.} \bibnamefont{Hamann}},
  \bibnamefont{and}
  \bibinfo{author}{\bibfnamefont{M.}~\bibnamefont{Schl\"uter}},
  \bibinfo{journal}{Phys. Rev. B} \textbf{\bibinfo{volume}{26}},
  \bibinfo{pages}{4199} (\bibinfo{year}{1982}).

\bibitem[{\citenamefont{Kerker}(1980)}]{Kerker1980}
\bibinfo{author}{\bibfnamefont{G.}~\bibnamefont{Kerker}},
  \bibinfo{journal}{Journal of Physics C: Solid State Physics}
  \textbf{\bibinfo{volume}{13}}, \bibinfo{pages}{L189} (\bibinfo{year}{1980}).

\bibitem[{\citenamefont{Vanderbilt}(1985)}]{Vanderbilt1985}
\bibinfo{author}{\bibfnamefont{D.}~\bibnamefont{Vanderbilt}},
  \bibinfo{journal}{Phys. Rev. B} \textbf{\bibinfo{volume}{32}},
  \bibinfo{pages}{8412} (\bibinfo{year}{1985}).

\bibitem[{BLP()}]{BLPS}
\bibinfo{note}{Http://www.princeton.edu/carter/research/local-pseudopotentials/}.

\bibitem[{\citenamefont{Birch}(1947)}]{3B}
\bibinfo{author}{\bibfnamefont{F.}~\bibnamefont{Birch}},
  \bibinfo{journal}{Phys. Rev.} \textbf{\bibinfo{volume}{71}},
  \bibinfo{pages}{809} (\bibinfo{year}{1947}).

\bibitem[{\citenamefont{Wyckoff}(1963)}]{ex}
\bibinfo{author}{\bibfnamefont{R.}~\bibnamefont{Wyckoff}},
  \textbf{\bibinfo{volume}{1}}, \bibinfo{pages}{85} (\bibinfo{year}{1963}).

\end{thebibliography}

\end{document}